\documentclass[aps,prl,twocolumn,showpacs]{revtex4}

\usepackage[dvips]{graphicx}
\usepackage{amssymb}
\usepackage{amsmath}

\begin{document}

\title{Non-linear rheology of a nanoconfined simple fluid}

\author{Lionel Bureau}
 \email{bureau@insp.jussieu.fr}
\affiliation{Institut des Nanosciences de Paris, UMR 7588 CNRS-Universit\'e Paris 6, 140 rue de Lourmel, 75015 Paris, France}

\date{\today}

\begin{abstract}
We probe the rheology of the model liquid octamethylcyclotetrasiloxane (OMCTS) confined into molecularly thin films, using a unique Surface Forces Apparatus allowing to explore a large range
of shear rates and confinement. We thus show that OMCTS under increasing confinement exhibits the viscosity enhancement and the non-linear flow properties characteristic of a sheared supercooled liquid
approaching its glass transition. Besides, we study the drainage of confined OMCTS via the propagation of ``squeeze-out'' fronts. The
hydrodynamic model proposed by Becker and Mugele [Phys. Rev. Lett. {\bf 91}, 166104 (2003)] to describe such front dynamics leads to a conclusion in apparent contradiction with the dynamical slowdown 
evidenced by rheology measurements, which suggests that front propagation is not controlled by large scale flow in the confined films.
\end{abstract}


\maketitle


Supercooled liquids share qualitative rheological features upon approaching the glass transition \cite{melt}: (i) their viscosity increases
dramatically, and (ii) they exhibit non-newtonian properties (shear thinning) when the time scale of 
mechanical forcing becomes shorter than that of structural relaxation. The precise origin of such a behavior is the subject of active theoretical and numerical investigations \cite{theory}. 
Recently, an extension to flow situations of the mode-coupling 
theory (MCT) has been proposed, in order to describe this non-linear rheology \cite{FC}. Now, a stringent test of theoretical predictions against experimental results
requires measurements, over a large range of shear rate ($\dot{\gamma}$), of the non-linear properties as jamming is gradually approached. These are extremely challenging to perform on atomic glass formers, because of their elevated glass transition temperature
and flow stress. 
The most comprehensive studies to date have focused on colloidal suspensions of thermosensitive particles, in which the volume fraction, hence the distance to jamming, can be 
finely tuned \cite{ballauff}. It has thus been shown that, in very good agreement with MCT,
the flow stress of such suspensions exhibits a rate dependence all the weaker that the distance to glass 
transition is small, until a yield stress develops when the suspension gets jammed \cite{ballauff}. Such a behavior can be considered as the 
rheological hallmark of the approach to glass transition.

Here, we show that increasing confinement represents an alternative pathway to bring a system close to its jammed state under well-controlled conditions. Surface Forces Apparatus (SFA) experiments have shown 
that simple liquids confined between solid walls 
below thicknesses of a few molecular diameters exhibit enhanced flow resistance \cite{IMH}. From SFA experiments probing the linear response of ultrathin liquid films, Demirel and Granick (DG)
concluded to a confinement-induced dynamical slowdown, akin to what occurs in supercooled liquids \cite{DG}. However, this conclusion has been challenged by other groups probing the large strain shear response of 
confined fluids \cite{KK,YI}. Moreover, experiments by Becker and Mugele (BM) have shown that a confined liquid drains stepwise by expelling monolayers via the 
propagation of  ``squeeze-out fronts'' \cite{BM}. A model of the 
front dynamics, extending the work of Persson and Tosatti \cite{PT}, led them to conclude that the confined fluid retained its bulk viscosity, dissipation enhancement arising 
from high friction on the confining walls. 

The nature of the mechanisms by which the properties of liquids are affected by
confinement at the molecular scale therefore remains an open question. Such an issue, which is of interest for the fundamental understanding of the jamming transition \cite{weeks}, is also of paramount importance 
for boundary lubrication \cite{bo}, 
and for nanofluidics, where the knowledge of the flow properties of liquids confined into nanometer-sized channels or structures is crucial \cite{nanoflu}.

In this Letter, we report on the first SFA study in which {\it both} large strain shear rheology and squeeze-out fronts measurements are performed, in the same 
experimental run, on the nonpolar liquid octamethylcyclotetrasiloxane (OMCTS), which has been used in the aforementioned works. 

(i) We show unambiguously, from flow curves measured over 6 decades of $\dot{\gamma}$,  that OMCTS under increasing confinement exhibits the 
viscosity enhancement and non-newtonian features of a supercooled liquid
approaching the glass transition. 

(ii) We observe squeeze-out front dynamics in quantitative agreement with that previously reported \cite{BM}.  When analyzed within the framework of the BM model, it results in an effective viscosity
 two orders of magnitude lower than that directly measured in shear. We conclude that such an apparent contradiction arises from an improper assumption by BM about the nature of the 
 mass transport mechanism at play during  front propagation.

\begin{figure}[htbp]
$$
\includegraphics[width=8cm]{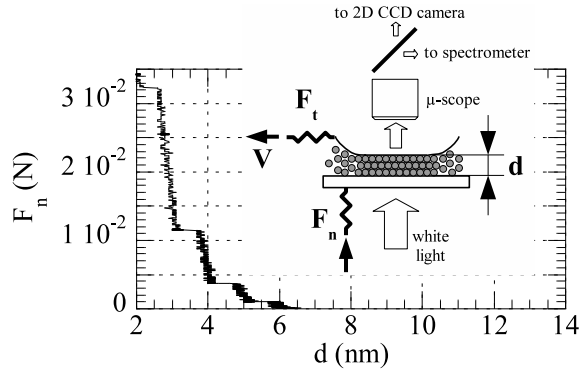}
$$
\caption{Force $vs$ distance curve during approach of the surfaces (loading velocity 0.5 nm.s$^{-1}$). Inset: scheme of the setup.  
White light is shone on the confined film, and the transmitted intensity is sent (i) to a spectrometer for spectral analysis \cite{IA}, 
and (ii) to a CCD camera acquiring images of the contact area $A$ at a rate of 55 s$^{-1}$.
}
\label{fig:fig1}
\end{figure}

Experiments were performed on a home-built SFA \cite{LB} (Fig. \ref{fig:fig1}). The liquid is confined between two atomically 
smooth backsilvered mica sheets glued
onto crossed cylindrical lenses (radius of curvature $\sim$1 cm). The normal force $F_{n}$ is measured by means of a
load cell of stiffness 9500 N.m$^{-1}$. The ``contact'' area $A$, over which the mica sheets elastically 
 flatten to form a circular parallel gap in which the liquid is confined, is monitored by videomicroscopy. The thickness of the film, $d$, is determined by multiple beam interferometry
 \cite{IA} and fast spectral correlation \cite{Heuberg1,LB}. Once confined under a given load, over an area $A$, the 
 liquid is sheared, in a plane Couette geometry, by moving laterally one surface at a velocity $V$ in the range 10$^{-4}-10^{2}\, \mu$m.s$^{-1}$, while measuring the resulting tangential force $F_{t}$ with a cell 
 of stiffness 5200 N.m$^{-1}$.  The shear stress sensitivity of the instrument is $F_{t}/A\sim 200$ Pa. Our SFA has the unique feature of using the normal force signal as the input of a feedback loop, which allows to 
 perform steady-state experiments over large shear amplitudes (up to hundreds
 of microns) under constant normal load conditions, whatever the level of confinement of the liquid.
The mica sheets were prepared as described in \cite{LB2}, glued onto the cylindrical lenses using a UV curing glue (NOA 81, Norland), and cleaved with adhesive tape immediately before being 
installed in the SFA, so as to obtain contaminant-free surfaces \cite{FS}. OMCTS from Fluka (purum grade $\geq$99\%) was vacuum distilled before use. A drop ($\sim 150 \, \mu$L) 
of the liquid, filtered through a 0.2 $\mu$m membrane, was injected between the surfaces. It was then left at $T=20\pm 0.01^{\circ}$C for 12h in the sealed SFA, containing P$_{2}$O$_{5}$ 
to scavenge residual moisture, before beginning experiments.

Fig. \ref{fig:fig1} shows a force-distance profile measured upon quasi-static approach of the surfaces: it is clearly seen that below 6 nm\cite{footnote1}, the thickness of the confined liquid decreases by steps 
of approximately 8 \AA, which corresponds to the minor diameter of the slightly oblate OMCTS molecule. This reflects the well-documented
 wall-induced layered structure of the fluid, which gives rise to the so-called solvation forces \cite{solvat}.
  
\begin{figure}[htbp]
$$
\includegraphics[width=8cm]{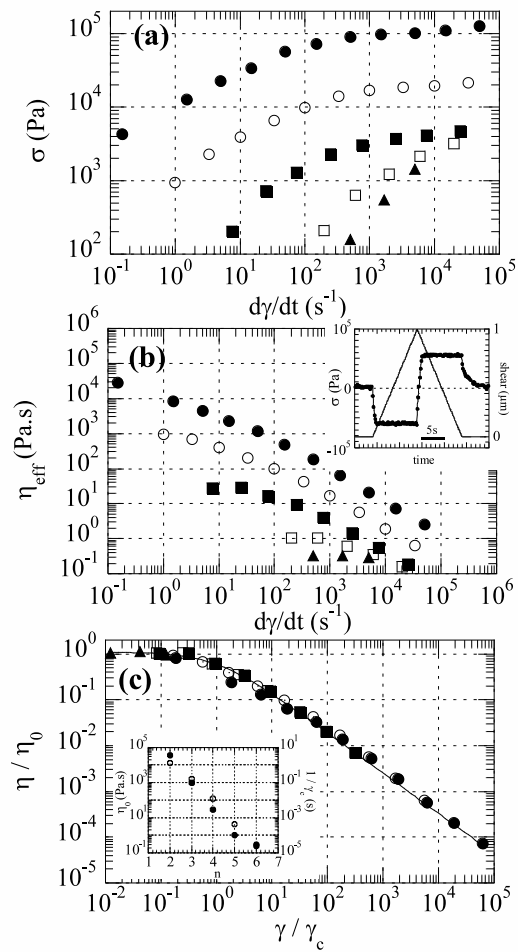}
$$
\caption{(a) $\sigma(\dot{\gamma})$ for OMCTS films of ($\blacktriangle$) 6, ($\square$) 5, ($\blacksquare$) 4, ({\large $\circ$}) 3, and ({\large $\bullet$}) 2 monolayers. (b)
$\eta_{\text{eff}}(\dot{\gamma})$, symbols as in (a). Insert: time trace of ({$\bullet$}, left scale) $\sigma$ measured at $V=0.1\,\mu$m.s$^{-1}$ on a 2nm-thick film, (line, right scale) the
forth and back shear motion applied. (c) Master curve showing data from (b) plotted as $\eta_{\text{eff}}/\eta_{0}$ $vs$ $\dot{\gamma}/\dot{\gamma}_{c}$. The solid line is a fit of the form
$\eta_{\text{eff}}/\eta_{0}= 1/(1+\dot{\gamma}/\dot{\gamma}_{c})^{0.88}$. Inset: values for $\eta_{0}$ ($\bullet$, left scale) and $1/\dot{\gamma}_{c}$ ($\circ$, right scale) used 
for each $n$.
}
\label{fig:fig2}
\end{figure}

We first focus on shear experiments performed on layered OMCTS films with thicknesses ranging from 6 down to 2 monolayers. Over the whole range of confinement and 
velocity explored, we have observed: (i) a smooth stable shear response (see time trace in the inset of Fig. \ref{fig:fig2}b), and (ii) a steady-state value of $F_{t}$ which increases with $V$.
On Fig. \ref{fig:fig2}a, we plot the steady-state flow stress $\sigma=F_{t}/A$ versus shear rate $\dot{\gamma}=V/d$ for the different film thicknesses. The same data are plotted on 
Fig. \ref{fig:fig2}b as the effective viscosity $\eta_{\text{eff}}=\sigma/\dot{\gamma}$ versus $\dot{\gamma}$. 

It can be seen that, as the OMCTS thickness is reduced from $n=$6 to 2 monolayers:

(i)  The flow stress, and hence the viscosity, steadily increases \cite{footnote2}. 

(ii) The dependence of $\sigma$ on $\dot{\gamma}$ shifts from linear (newtonian) to sublinear (shear-thinning). 

(iii)  The crossover shear rate, $\dot{\gamma}_{c}$, above which non-newtonian behavior is observed, shifts to smaller values.
 

(iv) For $n\leq$4 monolayers, power-law shear thinning ($\sigma\sim \dot{\gamma}^{\alpha}$, $\alpha<1$) at low $\dot{\gamma}$ crosses over to a quasi-plateau regime.

This set of features is characteristic of the approach to jamming, as reported experimentally \cite{ballauff} and predicted by MCT and numerical 
simulations \cite{theory,FC}.  This is further supported by the fact that $\eta_{\text{eff}}(\dot{\gamma})$ curves measured for different $n$ collapse onto a single master curve when plotted as $\eta_{\text{eff}}/\eta_{0}$ $vs$
$\dot{\gamma}/\dot{\gamma}_{c}$, with $\eta_{0}$ the zero shear viscosity (Fig. \ref{fig:fig2}c). The inset of Fig. \ref{fig:fig2}c
shows that both $\eta_{0}$ and $1/\dot{\gamma}_{c}$ ({\it i.e.} the relaxation time of the liquid) sharply increase as the film thickness is decreased. The reduced viscosity 
obeys $\eta_{\text{eff}}/\eta_{0}\simeq 1/(1+\dot{\gamma}/\dot{\gamma}_{c})^{0.88}$, which is consistent with theoretical predictions for sheared supercooled systems \cite{theory}.
Finally, the observation of a quasi-plateau regime which does not extend down to the lowest shear rates indicates that, in the present experiments, confined OMCTS 
approaches but does not reach jamming. This is consistent with the fact that (see inset of Fig. \ref{fig:fig2}b), upon cessation of shear, the stress relaxes (i) very slowly, over $\sim5$ s, and (ii) 
down to a non-measurable level. 

These observations lead us to conclude, in good agreement with DG \cite{DG}, that OMCTS undergoes dynamical slowdown upon increasing confinement, similarly to a 
supercooled system close above jamming. Such a conclusion contrasts
with that of Klein \cite{KK} or Israelachvili \cite{YI}, who observed responses exhibiting a stick-slip dynamics which they interpret in terms of shear-melting of a confinement-induced 
ordered solid-like structure. Such a discrepancy might have two origins. (i) The use of different protocols for mica surface preparation: Indeed, the method employed
in \cite{KK,YI}, in contrast to that described above, may lead to surface contamination by a submonolayer of nanoparticles, which have been suggested as a possible reason for the observed stick-slip behavior 
\cite{Pt}. (ii) Differences in the crystallographic alignment of the confining surfaces, which is expected to affect the shear response of the intercalated molecular film \cite{misfit}. No
systematic investigation have been made so far of the effect of alignment between contaminant-free surfaces, and we therefore cannot discriminate between point (i) and (ii) above to explain differences. 

\begin{figure}[htbp]
$$
\includegraphics[width=8cm]{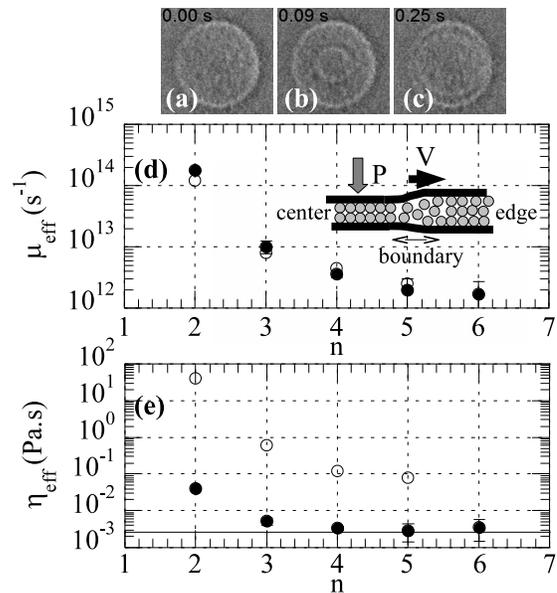}
$$
\caption{(a-c): Sequence of images (96$\times 96$ $\mu$m$^{2}$) showing the front propagation during a $3\rightarrow 2$ transition. 
(d): $\mu_{\text{eff}}$ vs $n$ (number of monolayers). ($\bullet$)  our results, ($\circ$) BM results, adapted from \cite{BM}. Inset: schematic cross-section of the film during squeeze-out. 
(e): $\eta_{\text{eff}}$ vs $n$. ($\circ$) measured in shear, and ($\bullet$) deduced from squeeze-out experiments. The horizontal line indicates the bulk viscosity of OMCTS.}
\label{fig:fig3}
\end{figure}

We now present the results from squeeze-out experiments. During loading, we record the light intensity transmitted through the contact
area, along with $F_{n}$ and $d$. We observe, as in \cite{BM,LB2}, that a film of thickness $n$ monolayers drains via nucleation/growth of a circular region of 
thickness $(n-1)$ layers (see Fig. \ref{fig:fig3}). Nucleation is accompanied by elastic relaxation of the confining sheets, which are locally bent in the boundary zone connecting the regions of thickness
 $n-1$ and $n$ (Fig. \ref{fig:fig3} inset). This 
creates a 2D pressure gradient which then drives the monolayer expulsion \cite{BM}. The local curvature of the mica sheets induces a contrast in the transmitted intensity (Fig. \ref{fig:fig3}a-c)
which allows us to follow with time the position of the ``squeeze-out'' front. We have thus measured, for successive $n\rightarrow n-1$ transitions,
 the squeeze-out time $\tau$ needed to expel one monolayer from the contact area $A$. We have done so before and after rheology experiments, and did not observe any
influence of shear history on front dynamics.

In the Persson and Tosatti (PT) model \cite{PT}, the front velocity is related to the 2D pressure gradient by: $\nabla p_{\text{2D}}=-\rho_{\text{2D}}\mu_{\text{eff}}V$,
where $p_{\text{2D}}\sim Pa$ and $\rho_{\text{2D}}=\rho a$ \cite{PT,BM} ($P=F_{n}/A$ is the applied pressure, $\rho$  the fluid density and $a$ the molecular size), $V$ is the front 
velocity and $\mu_{\text{eff}}$ a viscous drag coefficient. The latter 
is deduced from the squeeze-out time as \cite{PT}: $\mu_{\text{eff}}=4\pi\tau P/(\rho A)$. On Fig. \ref{fig:fig3}d we have plotted $\mu_{\text{eff}}$ as a function of film thickness for our 
experiments, along with the values obtained by Becker and Mugele (BM) \cite{BM}. There is quantitative agreement between both data sets. BM have extended the PT model by assuming that
front propagation is controlled by a ``layered'' Poiseuille flow between the front and the edge of the confinement area (Fig. \ref{fig:fig3} inset), and thus proposed that $\mu_{\text{eff}}$ should identify with 
the drag coefficient of a Hele-Shaw flow, {\it i.e.} 
$\mu_{\text{eff}}= 12 \tilde{\eta}_{\text{eff}}/(\rho d^{2})$, with $\tilde{\eta}_{\text{eff}}$ a shear viscosity and $d$ the film thickness \cite{BM}. We use this expression to infer 
$\tilde{\eta}_{\text{eff}}(d)$ from the front 
dynamics. On Fig. \ref{fig:fig3}e, we compare \cite{footnote3} it to $\eta_{\text{eff}}(d)$ obtained from shear data. It appears that $\tilde{\eta}_{\text{eff}}$,
which stays close to the bulk value down to 3-layer-thick films, is about two orders of magnitude lower than $\eta_{\text{eff}}$. 

We propose the following explanation to this apparent paradox.
Shear experiments are a straightforward way to measure $\eta_{\text{eff}}$, in contrast to squeeze-out experiments, which require modelling of the front dynamics to infer a viscosity. 
Therefore, we consider that the reliable results regarding $\eta_{\text{eff}}(d)$ are those from shear rheology. We are then left with the observation of squeeze-out
fronts which, given the $\eta_{\text{eff}}(d)$ obtained in shear, travel much faster than expected from the drag mechanism assumed in the BM picture, from which we have also drawn erroneous conclusions in a 
recent study \cite{LB2}. 
This suggests that the front dynamics is not controlled by the coherent sliding of adjacent incompressible molecular layers ahead of the front.
Indeed, another piece of information emerges from the force-distance 
profile of Fig. \ref{fig:fig1}: between two steps, the film thickness is observed to decrease by about 3 \AA~as the force is increased. Such a thickness variation is reversible upon load reduction.
 This shows that layered OMCTS films are substantially compressible, hence contain a non-negligible amount of free volume, which is consistent with the fact that confined films do not reach jamming.  
This certainly facilitates local rearrangements, and it is therefore likely that during propagation of a squeeze-out front, molecules in the layered region ahead of it 
permeate between layers in order to accommodate for density variations in the vicinity of the front. The apparent low resistance to front propagation suggests that permeation, rather than large scale coherent 
sliding of layers, controls mass transport ahead of the fronts. It implies that, pending further modelling, front dynamics cannot be used to infer a viscosity.

In summary, we have probed the rheology of a simple fluid under molecular confinement, and conclude that its behavior is akin to that of a sheared supercooled liquid close above the glass transition. 
This shows, as suggested by recent experiments on colloids \cite{weeks}, that confinement can be used as an alternative route to finely control the approach to jamming. 
Our results now raise two important questions. (i) We observe a liquidlike behavior down to the thinnest film investigated, which brings up the issue of how to cross the jamming transition under confinement. 
Two routes can be envisaged. It can be done by varying the chemical corrugation of the walls, as shown in friction experiments \cite{LB3} 
or in numerical simulations \cite{ayapa}, or, as mentioned above, by changing the orientation between the crystalline lattices of the confining surfaces. (ii) We find that 6 layer-thick films already
exhibit a viscosity two orders of magnitude larger than the bulk value. This raises the question of the scale below which non-bulk behavior appears, and how it compares to the range of surface forces.

We thank A. N. Morozov for fruitful discussion and C. Caroli for critical reading of the manuscript.



\begin{thebibliography}{}
\bibitem{footnote1} Steps are observed below $\sim$9 nm. The first ones, occuring at $F_{n}\leq$ 10$^{-5}$N, are
not visible on the plot of Fig. \ref{fig:fig1}.
\bibitem{footnote2} $\sigma (\dot{\gamma})$ have been measured under a pressure $p$ going from 1.5 MPa ($n=6$) to 7.5 MPa ($n=2$). We have checked that $\sigma$ depends  
weakly on $p$ at constant thickness. Its increase thus cannot be ascribe to an increase in $p$, and is due to geometrical confinement.
\bibitem{footnote3} We compare, for each $d$, $\tilde{\eta_{\text{eff}}}$ with $\eta_{\text{eff}}$ measured at a shear velocity $V$ equal to the average front velocity.
\end{thebibliography}
\end{document}